\documentclass[sigconf]{acmart}
\usepackage{lipsum}

\setcopyright{acmlicensed}
\copyrightyear{2026}
\acmYear{2026}

\renewcommand\footnotetextcopyrightpermission[1]{}
\acmConference{}{}{}

\usepackage[main=UKenglish]{babel}
\usepackage{xspace}
\usepackage[%
  group-minimum-digits=4,%
  list-final-separator={, and },%
  print-zero-integer=true,%
  free-standing-units,%
  round-mode=figures,%
  round-precision=3,%
  mode=match,%
  reset-text-series=false,%
  text-series-to-math=true,%
  uncertainty-mode=separate,%
  round-pad=false,
]{siunitx}
\usepackage[]{subcaption}
\usepackage{listings}
\lstdefinestyle{inlinelst}{%
  basicstyle=\normalsize\ttfamily,%
  keywordstyle=,%
  language=Python,%
}
\lstdefinestyle{numberedlst}{%
  language=Python,%
  basicstyle=\scriptsize\ttfamily,%
  numbers=left,%
  numberstyle=\tiny,%
  tabsize=2,%
  breaklines=true,%
  xleftmargin=1.9em,%
  xrightmargin=1em,%
}
\lstdefinestyle{nonnumberedlst}{%
  language=Python,%
  basicstyle=\scriptsize\ttfamily,%
  numbers=none,%
  tabsize=2,%
  breaklines=true,%
  xleftmargin=1em,%
  xrightmargin=1em,%
}
\lstdefinestyle{nonnumberedtxt}{%
  basicstyle=\scriptsize\ttfamily,%
  numbers=none,%
  tabsize=2,%
  breaklines=true,%
  xleftmargin=1em,%
  xrightmargin=1em,%
}

\usepackage{paralist}
\usepackage{balance}
\usepackage{colortbl}
\usepackage{adjustbox}
\usepackage{multirow}
\usepackage{todonotes}
\usepackage{graphicx}
\usepackage{csquotes}
\usepackage{mismath}
\usepackage{tabularx}
\usepackage{booktabs}
\usepackage{threeparttable}
\SetExtraKerning{
  encoding = {T1}
}{
  \textemdash = {167,167},
  — = {167,167}
}
\usepackage[capitalise]{cleveref}
\usepackage{tikz}
\usetikzlibrary{arrows.meta, positioning, shapes.multipart,
shapes.geometric, calc, backgrounds}

\usepackage{enumitem}
\newlist{rqlist}{enumerate}{1}
\setlist[rqlist,1]{label=\textbf{RQ\arabic*:}, ref=RQ\arabic*, nosep}

\usepackage{tcolorbox}
\usepackage{xspace}
\usepackage{todonotes}

\begin{document}

\title{Why LLMs Fail: A Failure Analysis and Partial Success Measurement for Automated Security Patch Generation}

\author{Amir Al-Maamari}
\email{almaam03@ads.uni-passau.de}
\affiliation{
  \institution{University of Passau}
  \country{Germany}
}

\begin{abstract}
Large Language Models (LLMs) show promise for Automated Program Repair (APR), yet their effectiveness on security vulnerabilities remains poorly characterized. This study analyzes 319 LLM-generated security patches across 64 Java vulnerabilities from the Vul4J benchmark. Using tri-axis evaluation (compilation, security via PoV tests, functionality via test suites), the analysis reveals that only 24.8\% of patches achieve full correctness, while 51.4\% fail both security and functionality. The dominant failure mode is semantic misunderstanding: LLMs produce syntactically valid code but apply incorrect repair strategies. The proposed Security Repair Score (SRS) quantifies this gap, showing LLMs preserve functionality (mean 0.832) but struggle with security (mean 0.251). Vulnerability type strongly predicts difficulty, with fix rates ranging from 0\% (input validation) to 45\% (infinite loop). These findings demonstrate that LLM security patches require rigorous validation before deployment.
\end{abstract}


\maketitle

\section{Introduction}

LLM-powered Automated Program Repair (APR) has achieved notable results on functional bug benchmarks such as Defects4J and SWE-bench~\cite{Yang2025ASO}. However, the security domain presents distinct challenges: developer test suites verify expected behavior but do not defend against adversarial inputs. A patch that passes all tests may still leave the system vulnerable.

Recent work confirms this concern. LLM agents introduce vulnerabilities at nearly nine times the rate of human developers~\cite{sajadi2025}, and security-hardening techniques frequently destroy functionality~\cite{dai2025}. These findings suggest a fundamental tension between security and functionality in LLM-generated code.

This study investigates LLM security patch generation through systematic failure analysis. Using 319 patches generated by Gemini 2.0 Flash across 64 Vul4J vulnerabilities, the analysis characterizes failure modes, quantifies partial success via continuous metrics, and identifies difficulty predictors. The central finding is that semantic misunderstanding dominates: over half of patches apply fundamentally incorrect repair strategies.

The contributions include: (1) a failure taxonomy for LLM security patches, (2) the Security Repair Score (SRS) for measuring partial success, (3) identification of CWE-specific difficulty patterns, and (4) actionable guidance for practitioners.
\section{Background}

Traditional APR evaluates patches using functional test suites: a patch is ``plausible'' if it passes all tests. Security vulnerabilities present unique challenges: test suites verify expected behavior, not security properties; security correctness is often binary (exploitable or not); and security failures are silent (correct outputs while remaining exploitable).

Proof-of-Vulnerability (PoV) tests~\cite{vul4j2022} address these challenges by providing exploit code that fails on vulnerable systems and passes on patched ones. Semgrep~\cite{semgrep} is an open-source static analysis tool that uses pattern-based rules to detect security vulnerabilities without executing code. This study uses Semgrep's ``security-audit'' ruleset, which contains curated rules for common vulnerability patterns. The combination of dynamic (PoV) and static (Semgrep) analysis provides defense-in-depth: PoV tests catch exploitable vulnerabilities while static analysis identifies suspicious patterns that may indicate incomplete fixes.

Lizard~\cite{lizard} is a code complexity analyzer that computes metrics including lines of code (LOC) and cyclomatic complexity. Cyclomatic complexity measures the number of linearly independent paths through code, serving as a proxy for code difficulty. This study uses Lizard to extract vulnerability-level features for correlation analysis.

To identify difficulty predictors, this study employs both Pearson and Spearman correlation coefficients \cite{Schober2018Correlation}. Pearson measures linear relationships and assumes normally distributed data, while Spearman measures monotonic relationships without distributional assumptions. Using both methods reveals whether relationships are linear (both significant) or monotonic but non-linear (only Spearman significant), providing richer insight into how vulnerability characteristics affect repair difficulty.

\section{Approach}

\begin{figure*}[t]
  \centering
  \includegraphics[width=\textwidth]{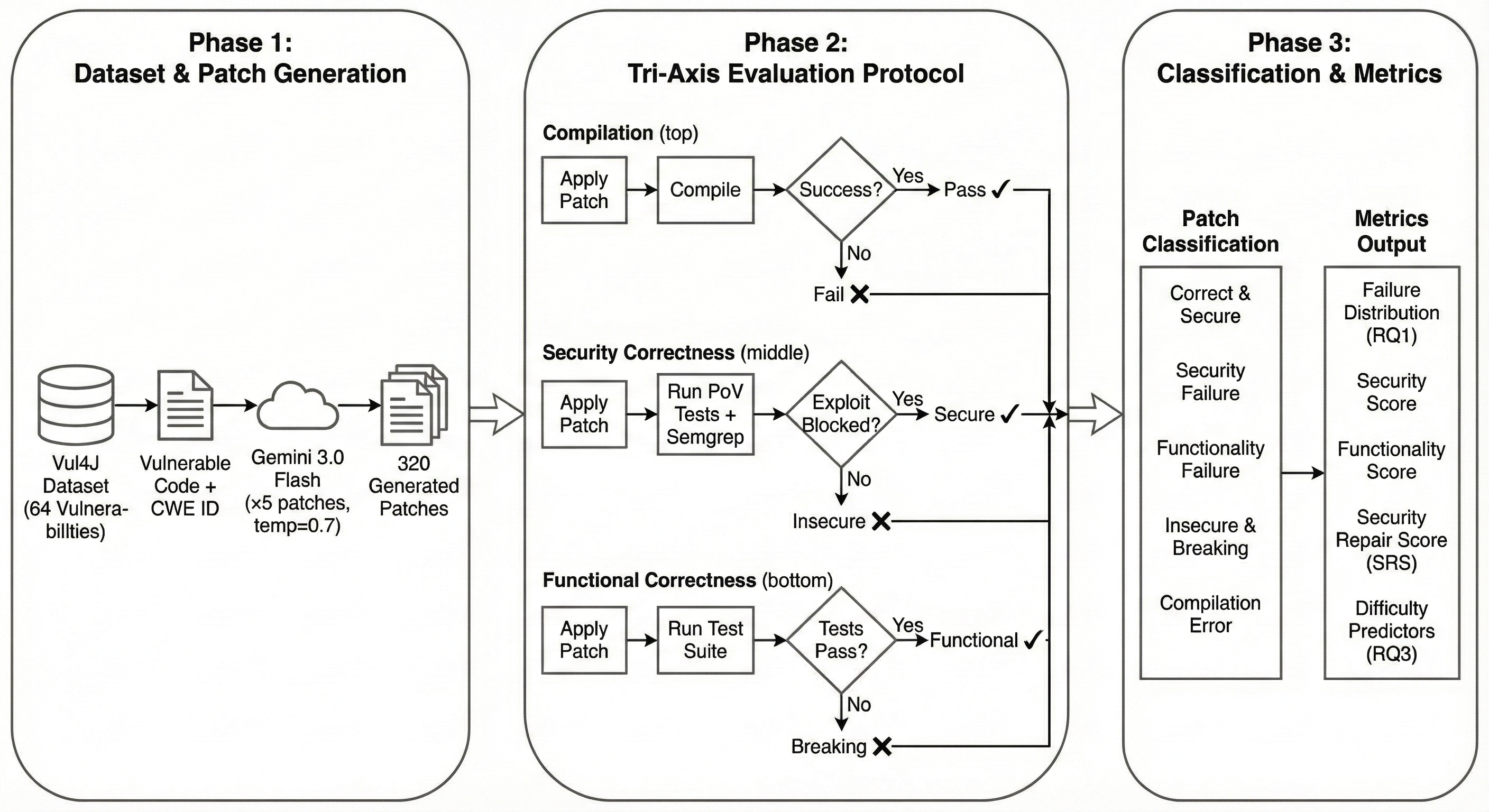}
  \caption{Experimental pipeline: vulnerable code $\rightarrow$ LLM $\rightarrow$ patch extraction $\rightarrow$ tri-axis evaluation.}
  \label{fig:pipeline}
\end{figure*}

This study uses an LLM with a zero-shot security repair prompt to generate patches for Java vulnerabilities. Figure~\ref{fig:pipeline} illustrates the pipeline. The following research questions guide the analysis:

\textbf{RQ1:} \textit{How do LLM-generated security patches fail?} This question categorizes patches into discrete failure modes to identify what goes wrong.

\textbf{RQ2:} \textit{To what degree do patches partially succeed?} This question uses continuous metrics to measure how much progress patches make toward correctness.

\textbf{RQ3:} \textit{Which vulnerability characteristics predict repair difficulty?} This question examines whether CWE type, complexity, or patch size correlate with mean SRS (partial success) across vulnerabilities.

\subsection{Evaluation Protocol}

Each patch is evaluated on three axes:
\begin{itemize}[noitemsep]
\item \textbf{Axis 1 (Compilation):} Project compiled using Maven/Gradle.
\item \textbf{Axis 2 (Security):} PoV test executed; Semgrep run for residual issues.
\item \textbf{Axis 3 (Functionality):} Full test suite executed.
\end{itemize}

Based on these axes, patches are classified into five categories: \textit{Correct \& Secure} (all pass), \textit{Compilation Error} (Axis 1 fails), \textit{Security Failure} (functional but insecure), \textit{Functionality Failure} (secure but breaks tests), and \textit{Insecure \& Breaking} (both Axes 2 and 3 fail).

\subsection{Metrics}

\textbf{Security Score} ($S_{score}$): PoV test result with Semgrep warning reduction:
\begin{equation}
S_{score} = P \times \left(1 - \min\left(\frac{W_{after}}{\max(W_{before}, 1)}, 1\right)\right)
\end{equation}
where $P = 1$ if PoV passes, else 0.

\textbf{Functionality Score} ($F_{score}$): $F_{score} = T_{passed} / T_{human}$

\textbf{Security Repair Score (SRS)}:
\begin{equation}
\text{SRS} = C \times (0.5 \cdot S_{score} + 0.5 \cdot F_{score})
\end{equation}
where $C = 1$ if patch compiles, else 0. SRS ranges from 0 to 1.
\section{Experimental Setup}

A replication package is available at: \url{https://github.com/AmeerAlmaamari/LLM-Security-Patches-Analysis}

\subsection{Dataset}

This study uses Vul4J~\cite{vul4j2022}, a benchmark of 79 Java security vulnerabilities with Docker-based reproduction. Each entry includes vulnerable code, human patch, PoV test(s), and developer test suite. Of 79 vulnerabilities, 64 were reproducible and included. The dataset spans 21 CWE categories.

\subsection{LLM}

Gemini 3.0 Flash~\cite{deepmind2025gemini3flash} generates patches using zero-shot prompting. Five patches per vulnerability (temperature 0.7) yield 320 patches; one excluded due to infrastructure timeout, leaving 319 for evaluation.

\begin{tcolorbox}[colback=white, colframe=black, title={\textbf{Prompt Template}}, fonttitle=\small]
\small
\texttt{You are a security expert. The following Java code contains [CWE-ID] ([CWE-Name]).}

\texttt{Task: Fix the security vulnerability. Preserve all original functionality.}

\texttt{Requirements:}
\begin{itemize}[noitemsep,topsep=2pt,leftmargin=*]
\item \texttt{Return ONLY the complete corrected Java code}
\item \texttt{No explanations, comments, or markdown formatting}
\item \texttt{Maintain the exact same class/method structure}
\end{itemize}
\end{tcolorbox}

This minimal prompt tests the model's intrinsic security knowledge rather than its ability to follow detailed remediation instructions. The LLM response is parsed to extract Java code: if the response contains markdown code blocks, the content within the blocks is extracted; otherwise, the entire response is treated as code.

\subsection{Threats to Validity}

\textit{Internal:} LLM non-determinism mitigated by 5 patches per vulnerability. Docker infrastructure may introduce confounds through environment-specific build failures or test flakiness; this was validated by confirming all 64 human patches compile and pass all tests in the same environment.

\textit{External:} Results specific to Java/Vul4J and Gemini 3.0 Flash. The 64-vulnerability sample covers 21 CWEs but may not represent all security bugs.

\textit{Data Contamination:} Low success rate (24.8\%) and 0\% fix rate for some CWEs suggest memorization is not driving results.
\section{Results and Discussion}

\subsection{RQ1: How Do Patches Fail?}

\begin{table}[t]
\centering
\caption{Patch outcome distribution (n=319).}
\label{tab:failure_dist}
\resizebox{\linewidth}{!}{%
\begin{tabular}{lrrl}
\toprule
\textbf{Category} & \textbf{Count} & \textbf{\%} & \textbf{Definition} \\
\midrule
Correct \& Secure & 79 & 24.8 & All axes pass \\
Insecure \& Breaking & 164 & 51.4 & Security and functionality fail \\
Compilation Error & 42 & 13.2 & Compilation fails \\
Security Failure & 33 & 10.3 & Functional but insecure \\
Functionality Failure & 1 & 0.3 & Secure but breaks tests \\
\bottomrule
\end{tabular}%
}
\end{table}

Table~\ref{tab:failure_dist} shows patch outcomes. Only 24.8\% achieve full correctness. The dominant failure is \textit{Insecure \& Breaking} (51.4\%): patches that neither fix the vulnerability nor preserve functionality. Subcategory analysis reveals that every non-compilation failure involves a wrong repair strategy: 143 patches (44.8\%) apply an incorrect strategy that also changes program behavior, 17 (5.3\%) violate API contracts, and 4 (1.3\%) overcorrect by removing too much functionality. This uniformity indicates a systematic gap in security reasoning, not random errors.

The compilation rate (86.8\%) indicates LLMs have learned Java syntax, yet syntactic correctness does not translate to security correctness. This disconnect is most visible in CWE-20 (Input Validation): 95\% of patches compile, but 0\% fix the vulnerability. Conversely, CWE-611 (XXE) has a lower compilation rate (80\%) but achieves 40\% fix rate, suggesting that compilation difficulty and security difficulty are orthogonal problems.

Pure security failures (10.3\%) pose the greatest deployment risk: these 33 patches pass all functional tests but remain exploitable, meaning they would pass CI/CD pipelines undetected. This category is disproportionately concentrated in CWE-264 (Permissions), where 35\% of patches are functional but insecure, compared to the overall 10.3\% rate. The implication is that access control vulnerabilities are particularly prone to producing deceptively ``correct'' patches.

\begin{tcolorbox}[colback=gray!10, colframe=gray!80, title={\textbf{Summary (RQ1: Semantic failures dominate)}}]
Only 24.8\% of patches are fully correct. Over half (51.4\%) fail both security and functionality due to fundamentally incorrect repair strategies, not syntax errors. The 10.3\% that are functional but insecure pose the greatest deployment risk.
\end{tcolorbox}

\subsection{RQ2: To What Degree Do Patches Partially Succeed?}

\begin{table}[t]
\centering
\caption{Continuous score statistics (n=319).}
\label{tab:scores}
\begin{tabular}{lcccc}
\toprule
\textbf{Metric} & \textbf{Mean} & \textbf{Median} & \textbf{Std} \\
\midrule
Security Score & 0.251 & 0.0 & 0.434 \\
Functionality Score & 0.832 & 0.999 & 0.364 \\
SRS & 0.542 & 0.499 & 0.318 \\
\bottomrule
\end{tabular}
\end{table}

Table~\ref{tab:scores} reveals a stark asymmetry between the two dimensions. Mean Functionality Score (0.832) exceeds Security Score (0.251) by 3.3$\times$. The interquartile ranges reinforce this gap: Functionality is tightly clustered at the top (IQR: 0.968--1.0), while Security spans the entire range (IQR: 0.0--1.0). This pattern indicates that LLMs reliably preserve existing program behavior but fail to address the specific security property under repair.

Critically, this asymmetry does not reflect a security-functionality tradeoff. Correlation analysis between Security Score and Functionality Score shows no significant relationship ($r=0.267$, $p>0.05$), meaning that fixing a vulnerability does not come at the cost of breaking functionality. The 79 perfect patches demonstrate that both objectives are simultaneously achievable; the remaining patches fail security for reasons unrelated to functionality preservation.

\begin{figure}[t]
\centering
\includegraphics[width=\columnwidth]{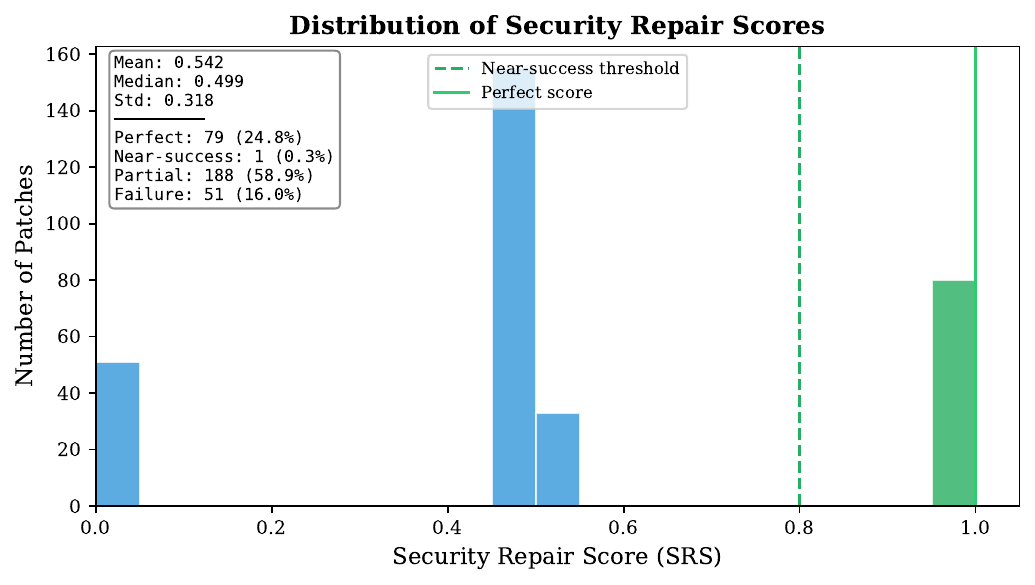}
\caption{SRS distribution showing bimodal pattern: clusters at SRS$\approx$1.0 (perfect) and SRS$\approx$0.5 (functional but insecure).}
\label{fig:srs_dist}
\end{figure}

\begin{table}[t]
\centering
\caption{Success tier distribution based on SRS.}
\label{tab:tiers}
\begin{tabular}{lrr}
\toprule
\textbf{Tier} & \textbf{Count} & \textbf{\%} \\
\midrule
Perfect (SRS=1.0) & 79 & 24.8 \\
Near-success (0.8$\leq$SRS$<$1.0) & 1 & 0.3 \\
Partial (0$<$SRS$<$0.8) & 188 & 58.9 \\
Complete failure (SRS=0) & 51 & 16.0 \\
\bottomrule
\end{tabular}
\end{table}

Figure~\ref{fig:srs_dist} shows a bimodal distribution. Patches cluster at SRS$\approx$1.0 (79 perfect) and SRS$\approx$0.5 (functional but insecure). Table~\ref{tab:tiers} quantifies this pattern: near-success patches (0.8$\leq$SRS$<$1.0) number only 1 (0.3\%), while 188 patches (58.9\%) achieve partial success but require significant work.

This bimodal pattern is not an artifact of the metrics: it emerges because Security Score is binary (PoV pass/fail) while Functionality Score is continuous. Patches that fail the PoV test receive $S_{score}=0$, yielding SRS$\approx$0.5 when functionality is preserved. The near-absence of near-success cases (0.3\%) carries a key insight: LLM security patching is an ``all-or-nothing'' capability. There is no gradient of partial security understanding that could be incrementally improved through prompt refinement or few-shot examples.

\begin{tcolorbox}[colback=gray!10, colframe=gray!80, title={\textbf{Summary (RQ2: Bimodal success pattern)}}]
LLMs preserve functionality (mean 0.832) but struggle with security (mean 0.251). The bimodal SRS distribution shows patches either succeed completely (24.8\%) or fail substantially, with almost no near-success cases (0.3\%). Minor adjustments to failed patches are unlikely to help.
\end{tcolorbox}

\subsection{RQ3: Which Characteristics Predict Difficulty?}

\begin{table}[t]
\centering
\caption{Correlation with mean SRS (n=64 vulnerabilities).}
\label{tab:correlations}
\resizebox{\linewidth}{!}{%
\begin{tabular}{lrrrr}
\toprule
\textbf{Feature} & \textbf{Pearson} & \textbf{$p$} & \textbf{Spearman} & \textbf{$p$} \\
\midrule
Lines of Code & $-0.179$ & 0.157 & $-0.041$ & 0.749 \\
Human Patch Size & $0.022$ & 0.863 & $-0.331$ & 0.008** \\
Cyclomatic Complexity & $0.107$ & 0.400 & $0.015$ & 0.907 \\
\bottomrule
\end{tabular}%
}
\end{table}

Table~\ref{tab:correlations} shows Human Patch Size has significant negative Spearman correlation ($\rho=-0.331$, $p=0.008$), while Pearson correlation is non-significant ($r=0.022$, $p=0.863$). This divergence indicates a monotonic but non-linear relationship: vulnerabilities requiring larger human patches are systematically harder for LLMs to repair, but the effect is not proportional. Traditional code complexity metrics (LOC, cyclomatic complexity) show no significant correlations. This is a meaningful negative result: it demonstrates that the challenge lies in understanding \textit{what} to change (semantic difficulty) rather than navigating complex code (structural difficulty).

\begin{table}[t]
\centering
\caption{CWE category analysis (well-sampled n$\geq$20). Only 6 of 21 CWE categories meet the n$\geq$20 threshold; remaining 139 patches span 15 categories with insufficient samples for reliable analysis.}
\label{tab:cwe_diff}
\begin{tabular}{lrrrrr}
\toprule
\textbf{CWE Category} & \textbf{n} & \textbf{Compile\%} & \textbf{SRS} & \textbf{Fix\%} \\
\midrule
CWE-835 (Infinite Loop) & 40 & 100 & 0.725 & 45 \\
CWE-20 (Input Validation) & 40 & 95 & 0.469 & 0 \\
CWE-79 (XSS) & 30 & 83 & 0.489 & 17 \\
CWE-611 (XXE) & 25 & 80 & 0.516 & 40 \\
CWE-22 (Path Traversal) & 25 & 72 & 0.458 & 20 \\
CWE-264 (Permissions) & 20 & 100 & 0.573 & 15 \\
\bottomrule
\end{tabular}
\end{table}

Table~\ref{tab:cwe_diff} reveals two distinct patterns. First, compilation rate and fix rate are disconnected: CWE-20 (Input Validation) compiles at 95\% but fixes at 0\%, while CWE-611 (XXE) compiles at only 80\% but fixes at 40\%. This confirms that the bottleneck is security reasoning, not code generation capability.

Second, the data reveals a spectrum from ``mechanical'' to ``semantic'' vulnerability types. CWE-835 (Infinite Loop) achieves 45\% fix rate with near-perfect functionality preservation (mean $F_{score}=0.999$), suggesting LLMs can identify loop termination conditions. CWE-20 (Input Validation) achieves 0\% fix rate despite high functionality ($F_{score}=0.938$), indicating that input validation requires domain-specific knowledge about what constitutes valid input in each application context. CWE-611 (XXE) shows a different pattern: lower functionality ($F_{score}=0.631$) but 40\% fix rate, suggesting that XXE patches often involve structural changes (disabling XML external entities) that affect both security and functionality simultaneously.

\begin{tcolorbox}[colback=gray!10, colframe=gray!80, title={\textbf{Summary (RQ3: Vulnerability type predicts difficulty)}}]
Fix rates range from 0\% (input validation) to 45\% (infinite loop). Human patch size correlates negatively with success ($\rho=-0.331$). Code complexity does not predict difficulty; vulnerability semantics does.
\end{tcolorbox}

\subsection{Implications}

\textbf{For Practitioners.} The 10.3\% of patches that are functional but insecure represent the most dangerous failure mode, as they would pass standard CI/CD testing and only be caught by security-specific validation such as PoV tests. This rate is not uniformly distributed: CWE-264 (Permissions) produces functional-but-insecure patches at 35\%, three times the overall rate. Teams should therefore prioritize security review by vulnerability type, focusing human effort on access control and input validation categories where LLM patches are systematically unreliable.

The bimodal success pattern (RQ2) has direct tooling implications: if an LLM-generated patch fails security validation, iterative prompt refinement is unlikely to help given that only 0.3\% of patches fall in the near-success range. The absence of a security-functionality tradeoff ($r=0.267$, $p>0.05$) means that teams should reject the assumption that fixing security requires accepting functionality regression.

\textbf{For Researchers.} The uniformity of wrong-strategy failures across all subcategories (RQ1) suggests that the core limitation is not syntax generation but vulnerability comprehension. Future work should investigate whether providing vulnerability-specific context (e.g., exploit descriptions, CWE remediation patterns) shifts the failure distribution from semantic to syntactic errors.

The CWE-specific difficulty spectrum (RQ3) points toward vulnerability type-specialized approaches. The 45\% fix rate for CWE-835 (mechanical fixes) versus 0\% for CWE-20 (semantic fixes) suggests that CWE-aware routing, directing different vulnerability types to different repair strategies, could substantially improve overall performance. The significant correlation between human patch size and difficulty ($\rho=-0.331$) further suggests that multi-step reasoning or chain-of-thought decomposition may help with vulnerabilities requiring complex patches.
\section{Related Work}

APR has evolved from heuristic-based to learning-based approaches using neural machine translation~\cite{xia2022}. Zero-shot learning with pre-trained code models can outperform fine-tuned models, and agentic frameworks like RepairAgent~\cite{bouzenia2025} achieve state-of-the-art results by autonomously localizing faults and refining patches. SAN2PATCH~\cite{san2patch} demonstrates that multi-stage reasoning with tree-structured prompting can achieve 79.5\% success on the VulnLoc dataset, decomposing patching into vulnerability comprehension, fault localization, fix strategy formulation, and patch generation. However, this body of work focuses on functional correctness or uses sanitizer logs rather than PoV tests for security validation.

Empirical studies quantify LLM security risks. Analysis of over 20,000 SWE-bench issues found LLMs introduce vulnerabilities at 9$\times$ the rate of human developers~\cite{sajadi2025}. SVEN~\cite{sven} addresses this through controlled code generation, using property-specific continuous vectors to guide generation toward secure code without modifying model weights, boosting secure code generation from 59.1\% to 92.3\%. Research on the tension between helpfulness and harmlessness shows that standard RLHF training causes models to prioritize task completion over safety~\cite{dai2025}.

SecurityEval~\cite{securityeval} provides 130 samples across 75 CWE-mapped vulnerability types for evaluating code generation models. 

CWEval~\cite{cweval} advances this with outcome-driven test oracles that assess both functionality and security simultaneously, revealing that existing benchmarks suffer from unclear specifications. Security-focused repair datasets include VUL4C~\cite{hu2025} (144 C/C++ vulnerabilities with exploits), PatchEval~\cite{wei2025} (1,000 CVEs across Go, JavaScript, and Python), and Vul4J~\cite{vul4j2022} (Java with PoV tests).

A methodological critique of existing work is the reliance on binary pass/fail metrics; ``plausible'' patches can serve as valid temporary mitigations~\cite{hu2025}. This study addresses this gap by contributing continuous metrics and a failure taxonomy that characterizes \textit{why} patches fail, not just whether they succeed.
\section{Conclusion}

This study analyzed 319 LLM-generated security patches across 64 Java vulnerabilities from the Vul4J benchmark, revealing three key findings. First, semantic misunderstanding dominates: over half of patches (51.4\%) apply fundamentally incorrect repair strategies, while compilation failures account for only 13.2\%. Second, LLMs preserve functionality (mean 0.832) substantially better than they fix security (mean 0.251), producing a bimodal success pattern with almost no near-misses. Third, vulnerability type strongly predicts difficulty, with fix rates ranging from 0\% (input validation) to 45\% (infinite loop) for well-sampled CWE categories.

The Security Repair Score (SRS) provides a continuous metric that captures partial progress, enabling meaningful evaluation even when full success is rare. For practitioners, these findings indicate that LLM-generated security patches require rigorous validation before deployment, with particular scrutiny for input validation vulnerabilities. For researchers, the CWE-specific difficulty patterns and the dominance of semantic failures suggest that security-aware training data and specialized reasoning capabilities represent the critical path forward.

\paragraph{Use of AI.} Gemini 2.5 Pro assisted with prose refinement. The author bears full responsibility for findings.

\bibliographystyle{ACM-Reference-Format}
\bibliography{library}

\end{document}